\documentclass[]{spie}  
\usepackage{graphicx}
 \DeclareGraphicsExtensions{.png,.pdf}
\usepackage{xparse}
\usepackage[detect-all]{siunitx}
\usepackage{booktabs}
\usepackage{booktabs}
\usepackage{multirow}
\usepackage{longtable}
\usepackage{comment}
\usepackage[colorlinks=true, allcolors=blue]{hyperref}

\title{Weighing Exo-Atmospheres: A novel mid-resolution spectral mode for SCALES}

\author[a,b]{R.~Deno Stelter}
\author[b]{Andrew~J. Skemer}
\author[a,b]{Renate Kupke}
\author[c]{Cyril Bourgenot}
\author[d]{Raquel~A. Martinez}
\author[d]{Stephanie~E. Sallum}
\affil[a]{UC Observatories, Santa Cruz, CA, USA}
\affil[b]{UC Santa Cruz, Santa Cruz, CA, USA}
\affil[c]{Durham University, Durham, UK}
\affil[d]{UC Irvine, Irvine, CA, USA}

\authorinfo{Corresponding author: R.D.S. E-mail: \href{mailto:deno@ucolick.org}{deno@ucolick.org}}

\begin{document} 
\maketitle

\begin{abstract}
SCALES (Slicer Combined with an Array of Lenslets for Exoplanet Spectroscopy) is a 2 - 5 micron high-contrast lenslet-based integral field spectrograph (IFS) designed to characterize exoplanets and their atmospheres. 
Like other lenslet-based IFSs, SCALES produces a short micro-spectrum of each lenslet’s micro-pupil. 
We have developed an image slicer that sits behind the lenslet array \& dissects and rearranges a subset of micro-pupils into a pseudo-slit. 
The combination lenslet array and slicer (or slenslit) allows SCALES to produce much longer spectra, thereby increasing the spectral resolution by over an order of magnitude and allowing for comparisons to atmospheric modeling at unprecedented resolution. 
This proceeding describes the design and performance of the slenslit.
\end{abstract}


\section{INTRODUCTION}
\label{sec:intro}  
\begin{figure}[htp]
    \centering
    \includegraphics[width=0.85\textwidth]{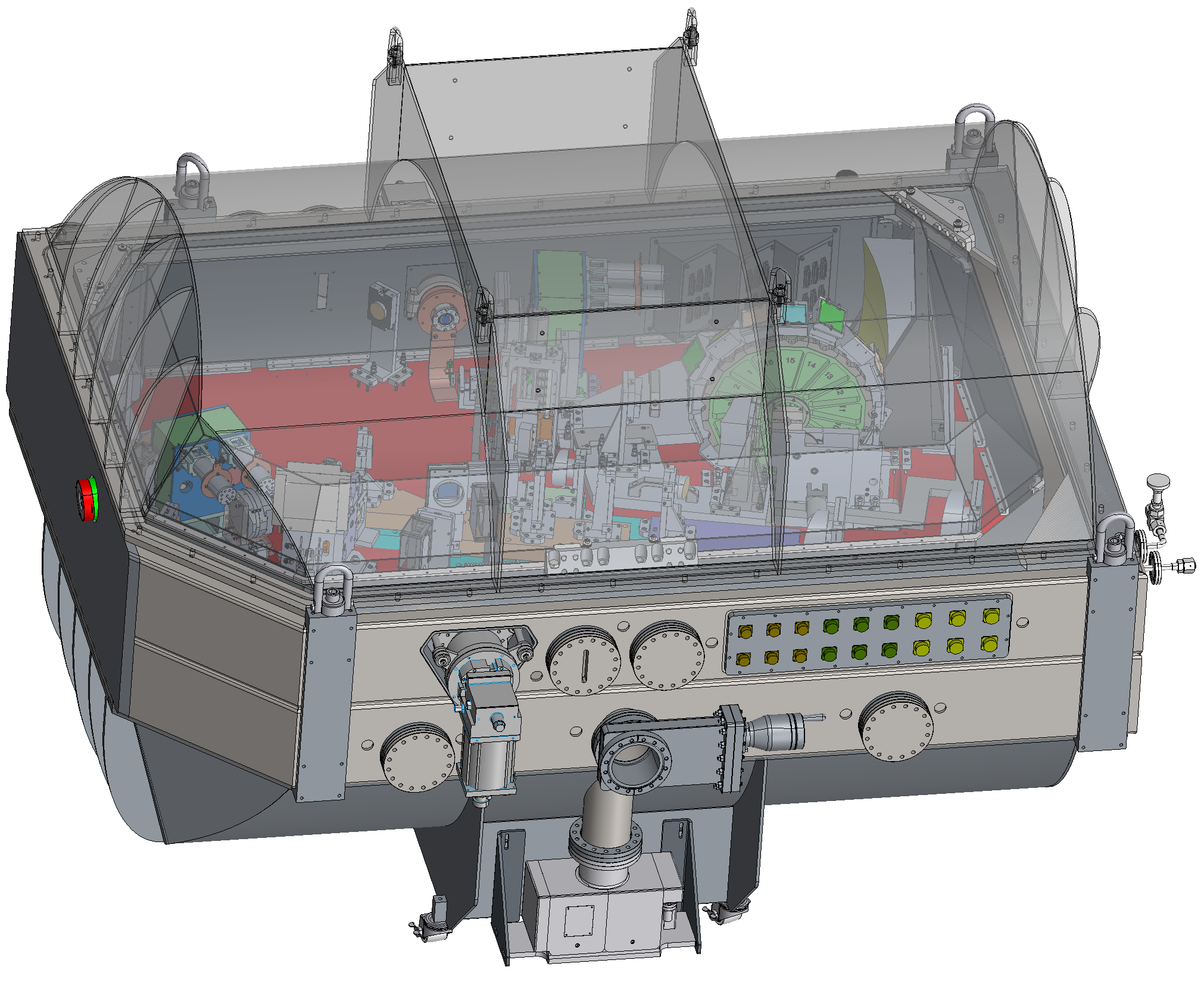}
    \caption{CAD of SCALES with the top lid and heat shield lid made transparent.
    }
    \label{fig:cad}
\end{figure}

Directly imaging exoplanets has driven the development of extreme AO (adaptive optics) and continues to push the technical envelope of astronomical instrumentation.
SCALES (\textsc{Slicer Combined with an Array of Lenslets for Exoplanet Spectroscopy}, previously the \textsc{Santa Cruz Array of Lenslets for Exoplanet Spectroscopy}), like previous direct imaging instruments such as GPI\cite{macintosh2014gpi} and CHARIS\cite{groff2015charis}, uses a lenslet array to sample the field of view and prisms to disperse each lenslet's pupil image into a short spectrum.
Figure~\ref{fig:cad} shows the a CAD model of SCALES.
The high-level specifications are shown in Table~\ref{tab:high-level-specs}.
It will sit behind the AO system of the Keck II telescope at Maunakea, where it will take full advantage of the large telescope primary diameter to peer closer in to stars hosting exoplanets than any other current OIR instrument.
SCALES is currently in its Final Design phase, and is planned to be on-sky by late 2025.

\begin{figure}[htp]
    \centering
    \includegraphics[width=0.85\textwidth]{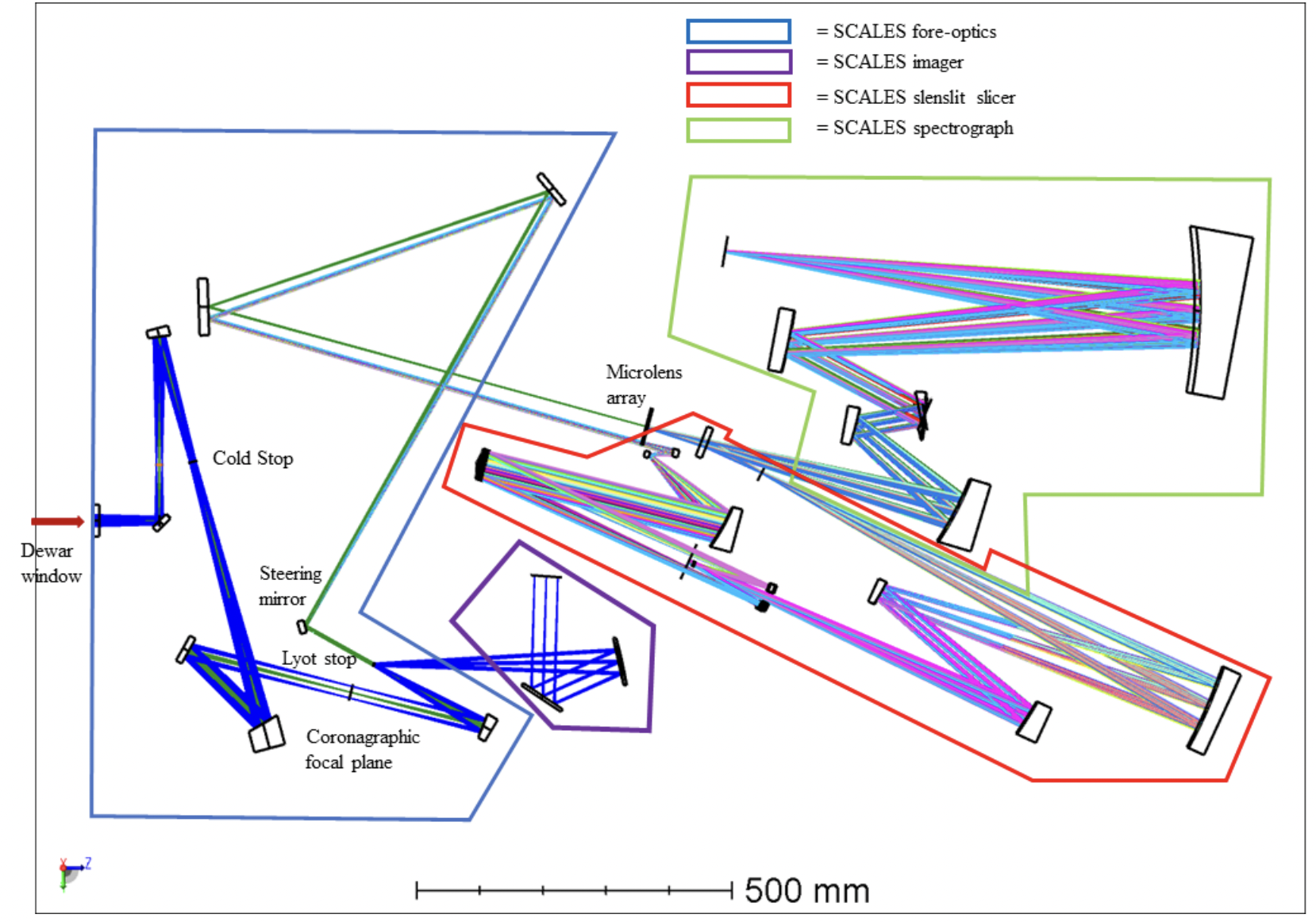}
    \caption{SCALES optical layout.
        The different boxes denote the SCALES modules.
        The slenslit is a `scenic bypass' for light on its way to the spectrograph.
    }
    \label{fig:optical-layout}
\end{figure}

Figure~\ref{fig:optical-layout} shows the SCALES optical design, with each module (foreoptics, imager, spectrograph, and slenslit) denoted with a polygon.
Coronagraphic optics, used to suppress starlight, are in the foreoptics and precede the lenslet array.
The spectrograph is a 1-1 reimaging system, and is next to the slenslit module which is used in the mid-resolution mode. 
It reimages the micro-pupil images produced by each lenslet array onto the detector, and selectable dispersers and filters are used to disperse the micro-pupil images into spectra onto the detector.

\begin{figure}[htp]
    \centering
    \includegraphics[width=0.85\textwidth]{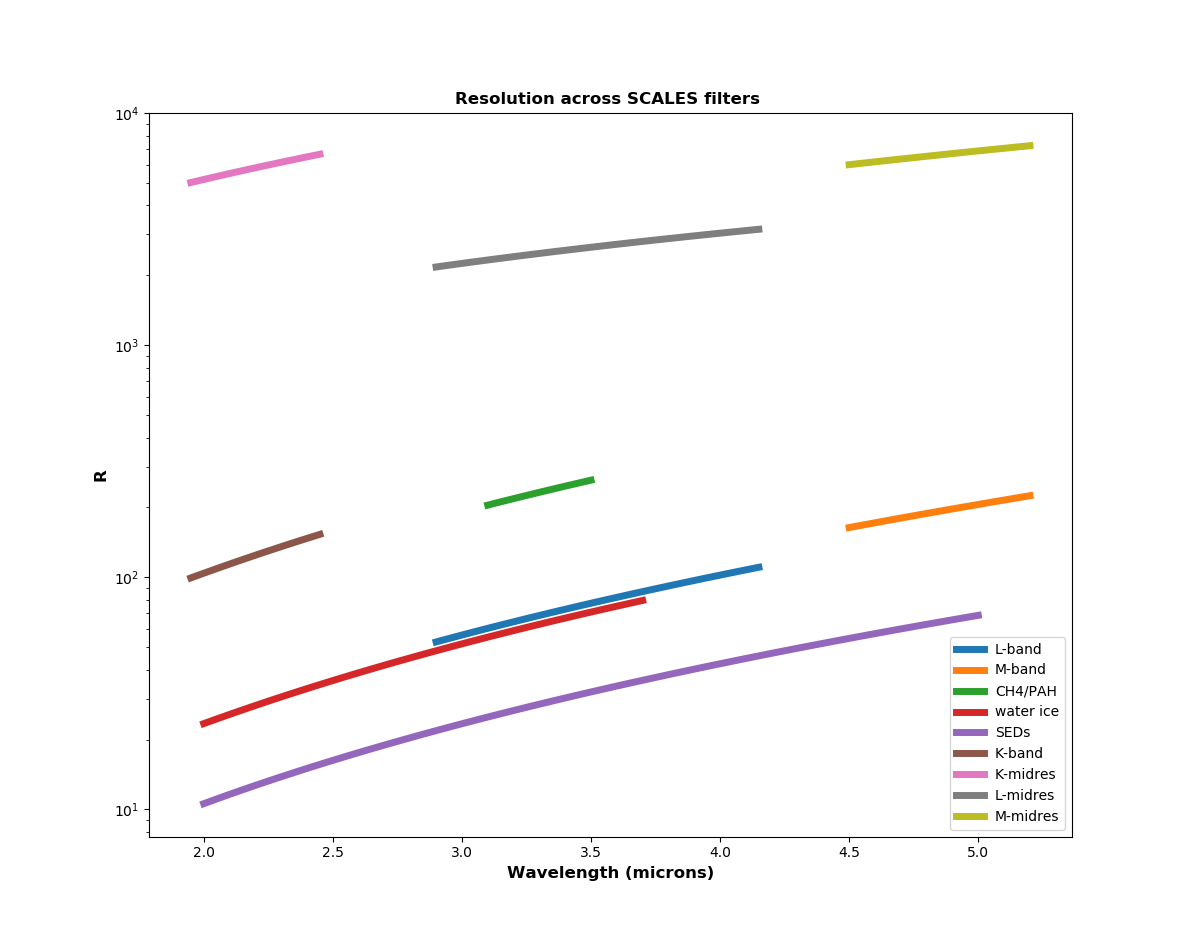}
    \caption{Spectral resolution for each bandpass, both low and mid-resolution.
        Spectral resolution is measured by looking at the instantaneous PSF at each wavelength for each mode.
        The slenslit offers more than a factor of 10 higher spectral resolution than the lenslet IFS does, while preserving Nyquist-limited spatial sampling across the 0.34'' x 0.36'' field of view.
    }
    \label{fig:resolution-by-filter}
\end{figure}

SCALES is designed with two main IFS modes: a low-resolution mode and a mid-resolution mode.
The low-resolution mode uses LiF prisms to disperse the spectra (see Table~\ref{tab-od:prism-specs} for details).
The mid-resolution mode uses custom 1st order gold-coated Zerodur gratings (see Table~\ref{tab-od:grating_specs}).
Figure~\ref{fig:resolution-by-filter} shows the instanteous spectral resolution as measured at the detector for each filter and mode combination.

The lenslet array has two sub-arrays; the low-resolution mode uses a 110x110 subarray, and a smaller 17x18 subarray used for the mid-resolution mode.
This mode will allow for characterization of exo-atmospheres at unprecedented spatial and spectral resolution at narrower inner working angles than ever before.
In doing so, we will be able to measure chemical abundances, in effect weighing exo-atmospheres.
There is also an imager mode, discussed in a separate Proceedings (see below).

\begin{table}[hpt]
    \centering
    \caption{SCALES high-level specs.} 

\begin{tabular}{|l|ll|ll|l|}
\hline
\multicolumn{1}{|c|}{\textbf{}}      & \multicolumn{2}{c|}{\textbf{Low-Resolution IFS}} & \multicolumn{2}{c|}{\textbf{Medium-Resolution IFS}}                                  & \multicolumn{1}{c|}{\textbf{Imager}}                                                            \\ \hline
\multirow{6}{*}{\textbf{Bandpass}} & \multicolumn{1}{l|}{2.0-2.4$\mu$m}  & R$\sim$150 & \multicolumn{1}{l|}{\multirow{2}{*}{2.0-2.4$\mu$m}}  & \multirow{2}{*}{R$\sim$6,000} & \multirow{6}{*}{\begin{tabular}[c]{@{}l@{}}Up to 16 filters \\ spanning 1-5$\mu$m\end{tabular}} \\ \cline{2-3}
                                     & \multicolumn{1}{l|}{2.0-4.0$\mu$m}  & R$\sim$50  & \multicolumn{1}{l|}{}                                &                               &                                                                                                 \\ \cline{2-5}
                                     & \multicolumn{1}{l|}{2.0-5.0$\mu$m}  & R$\sim$35  & \multicolumn{1}{l|}{\multirow{2}{*}{2.9-4.15$\mu$m}} & \multirow{2}{*}{R$\sim$3,000} &                                                                                                 \\ \cline{2-3}
                                     & \multicolumn{1}{l|}{2.9-4.15$\mu$m} & R$\sim$80  & \multicolumn{1}{l|}{}                                &                               &                                                                                                 \\ \cline{2-5}
                                     & \multicolumn{1}{l|}{3.1-3.5$\mu$m}  & R$\sim$200 & \multicolumn{1}{l|}{\multirow{2}{*}{4.5-5.2$\mu$m}}  & \multirow{2}{*}{R$\sim$7,000} &                                                                                                 \\ \cline{2-3}
                                     & \multicolumn{1}{l|}{4.5-5.2$\mu$m}  & R$\sim$200  & \multicolumn{1}{l|}{}                                &                               &                                                                                                 \\ \hline
\textbf{Field of View}               & \multicolumn{2}{l|}{2.15$\times$2.15"}           & \multicolumn{2}{l|}{0.36$\times$0.34"}                                               & 12.3$\times$12.3"                                                                               \\ \hline
\textbf{Spatial Sampling}            & \multicolumn{2}{l|}{0.02"}                       & \multicolumn{2}{l|}{0.02"}                                                           & 0.006"                                                                                          \\ \hline
\textbf{Coronagraphy}                & \multicolumn{2}{l|}{Vector-Vortex}               & \multicolumn{2}{l|}{Vector-Vortex}                                                   & TBD                                                                                             \\ \hline
\end{tabular}
    \label{tab:high-level-specs}
\end{table}

For more information on SCALES, please see the following proceedings in this conference:

\noindent\textbf{SCALES Overview}: Paper No. 12184-18 (Skemer et al.)\cite{SkemerStatus2022}\\
\textbf{Optical Design}: Paper No. 12184-159 (Renate Kupke et al.)\cite{Reni2022}\\
\textbf{Imaging Channel}: Paper No. 12188-65 (Ravinder Banyal et al.)\cite{Banyal2022}\\
\textbf{Cold-Stop / Lyot Stop}: Paper No. 12185-332 (Li et al.)\cite{Jialin2022}\\
\textbf{Aperture Masks}: Paper No. 12183-89 (Lach et al.)\cite{Lach2022}\\
\textbf{Keck Instrument Development}: Paper No. 12184-4 (Kassis et al.)\cite{Kassis2022}\\

\section{OPTICAL DESIGN OF THE SLENSLIT}
\label{sec:design}
The SCALES slenslit is responsible for reformatting a regular, two-dimensional grid of lenslet pupil image into a quasi-one-dimensional pseudoslit suitable for dispersing at medium spectral resolution.
It primarily uses three sets of mirrors to perform this rearrangement: 
\begin{enumerate}
    \item \textbf{Slicer mirrors} \\
        The slicer mirrors are responsible for separating each column of lenslet pupil images into an individual beam.
        The slices are located at a focal plane.
    \item \textbf{Pupil mirrors} \\
        The pupil mirrors, as the name suggests, are at the pupil planes of the slicer mirrors, and are not necessarily co-planar with each other.
        In combination with the slicer mirrors, the pupil mirrors de-magnify and place the images of the lenslet columns precisely where desired at the next focal plane (located at the field mirrors).
        However, given these powered optics' locations, they destroy the telecentricity of the outgoing beams.
    \item \textbf{Field mirrors} \\
        The field mirrors are at the output focal plane and co-planar. 
        By powering these optical surfaces, we can correct for the atelecentric errors introduced by the slicer and pupil mirrors.
        This allows us to keep the same size and location of the footprint on the disperser plane for all slices.
\end{enumerate}
In order to make the slicing optics fit inside the space envelope, we elected to use a set of input and output relays.
The input relay reimages the lenslet pupil image plane onto the slicer mirrors and magnifies the beam by a factor of 8, allowing the slicing optics operate at f/64.
The slicer performs a factor of 4 demagnification from the input focal plane at the slicer mirrors to the output focal plane at the field mirrors.  
The output relay returns the beam speed to f/8 to match the expected input speed to the collimator.
A return fold mirror that rides on the mode selector mechanism delivers the slenslit output beam into the spectrograph, which then disperses the pseudoslit onto the detector.
The mode selector mechanism carries vignetting masks and return fold mirror such that the low-resolution mode is used when the mid-resolution subarray of lenslets is blocked, and vice-versa.
The slenslit pseudoslit is confocal with the lenslet pupil image plane, meaning that the spectrograph `sees' both low and medium resolution inputs as originating at the same plane, and thus we can use the same optics and detector for both without refocusing.

\begin{table}[hpt]
    \centering
    \caption{Specifications of the SCALES prisms.} 

\begin{tabular}{p{4cm} | c p{2.25cm} p{2.5cm} p{2.5cm} }
\textbf{Wavelength range [\SI{}{\micro\meter}]} & \textbf{Mid-band spectral resolution} & \textbf{Apex Angle [deg]} & \textbf{Incident Angle [deg]}  \\
\midrule
\numrange{1.95}{2.45} (K-band) & R $\sim 150$ & 13.33 & 13.5    \\
\numrange{4.5}{5.2} (M-band) & R $\sim 70$ & 4.53 & 17.4 \\
\numrange{2.9}{4.15} (L-band) & R $\sim 40$  & 3.66  & 17.6   \\
\numrange{2.0}{5.0} (SEDS) & R $\sim 10$  & 1.54 & 18.4 \\
\numrange{3.1}{3.5} (CH\textsubscript{4}+PAH) & R $\sim 150$  & 11.46  & 14.5 \\
\numrange{2.0}{4.0} (water ice) & R $\sim 30$  & 3.36  & 18 \\
\bottomrule
\end{tabular}

    \label{tab-od:prism-specs}
\end{table}
\begin{table}[hpt]
    \centering
    \caption{Specifications of the SCALES gratings. The spectral resolution is calculated from linear dispersion, not the expected spectral resolution measured at the detector.} 

\begin{tabular}{p{5cm} | p{2cm} p{2cm} p{2cm}  p{2cm}}
\textbf{Wavelength range} & \textbf{Mid-band spectral resolution} & \textbf{Groove Density (lines/mm)}  & \textbf{Incident Angle (degrees)} & \textbf{Blaze Angle (degrees)} \\
\midrule
1.95 - 2.45 microns (K-band) & R = 4275 & 218 & 33.7 & 17.6  \\
2.9 - 4.15 microns (L-band) & R = 2716  & 86  & 28.2 & 10.1 \\
4.5 - 5.2 microns (M-band) & R = 6673 & 156 & 42.6 & 28.3 \\
\bottomrule
\end{tabular}
    \label{tab-od:grating_specs}
\end{table}

\subsection{Slenslit interleaving}
\label{subsec-od:slenslit-interleaving}
\begin{figure}[htp]
    \centering
    \includegraphics[width=0.5\textwidth]{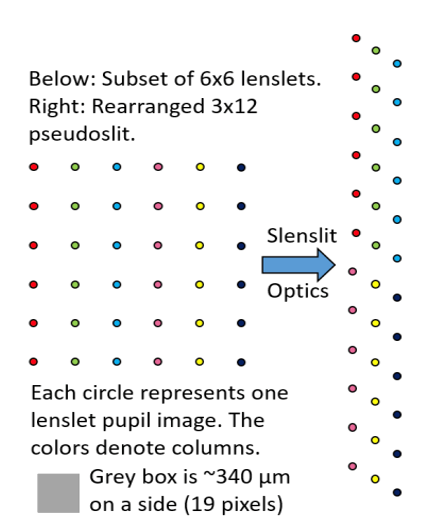}
    \caption[Interleaving of the slenslit pseudoslit.]
    {
        Interleaving schematic of the slenslit pseudoslit.
        The output super-columns are offset from one another in the vertical direction.
    }
    \label{fig-od:slenslit-interleaving}
\end{figure}
The interleaving of the pseudoslit comes about as a way to increase the effective field of view while preserving the spacing of each spectral trace on the detector relative to the low-resolution mode.
Figure~\ref{fig-od:slenslit-interleaving} shows a schematic using a $6 \times 6$ subarray of lenslet pupil images that are interleaved in the same way as the pseudoslit, although the SCALES pseudoslit is larger.
The SCALES slenslit pseudoslit is comprised of \SI{306}  pupil images (\SI{18} columns of \SI{17} rows each) arranged into 3 super-columns of \SI{6} columns of \SI{17} pupil images each. 
Each super-column is separated by \SI{1.5}{ \milli\meter} at the detector, which shortens the instantaneous bandpass slightly due to the spectra of the outer columns falling off of the detector.
Note that the separation of the spectra on the detector are separated by \numrange{5}{7} pixels on the detector, which is similar to the separation of the shorter spectra of the low-resolution mode.
The slenslit opto-mechanical design is discussed further in \S\ref{sec:mech-design}.

The slenslit optics are located after the lenslet array and feed the spectrograph with a rearranged pseudoslit for medium-resolution spectroscopy; the slenslit is the `scenic bypass' offering much higher spectral resolution over a smaller field of view compared to the low-resolution IFS.
Medium resolution dispersion is accomplished via a suite of three gratings, to be mounted on the same carousel and located at the same pupil plane as the prisms. 
We have baselined gold-coated etched gratings on Zerodur substrate with the characteristics shown in Table~\ref{tab-od:grating_specs}. 
The opening angle of \SI{38}{\degree} for all gratings and prisms is set by the geometry of the collimator and camera and the mechanical envelope needed for the large disperser carousel. 
The blaze angles in Table \ref{tab-od:grating_specs} were determined with GSolver, a rigorous coupled wave analysis program. These blaze angles give peak efficiencies of \SI{80}\% for the K- and L-bands and \SI{66}\% for M-band.

\subsection{Slenslit optical design}
\label{subsubsec-od:slenslit-design}
\begin{figure}[htp]
    \centering
    \includegraphics[width=0.95\textwidth]{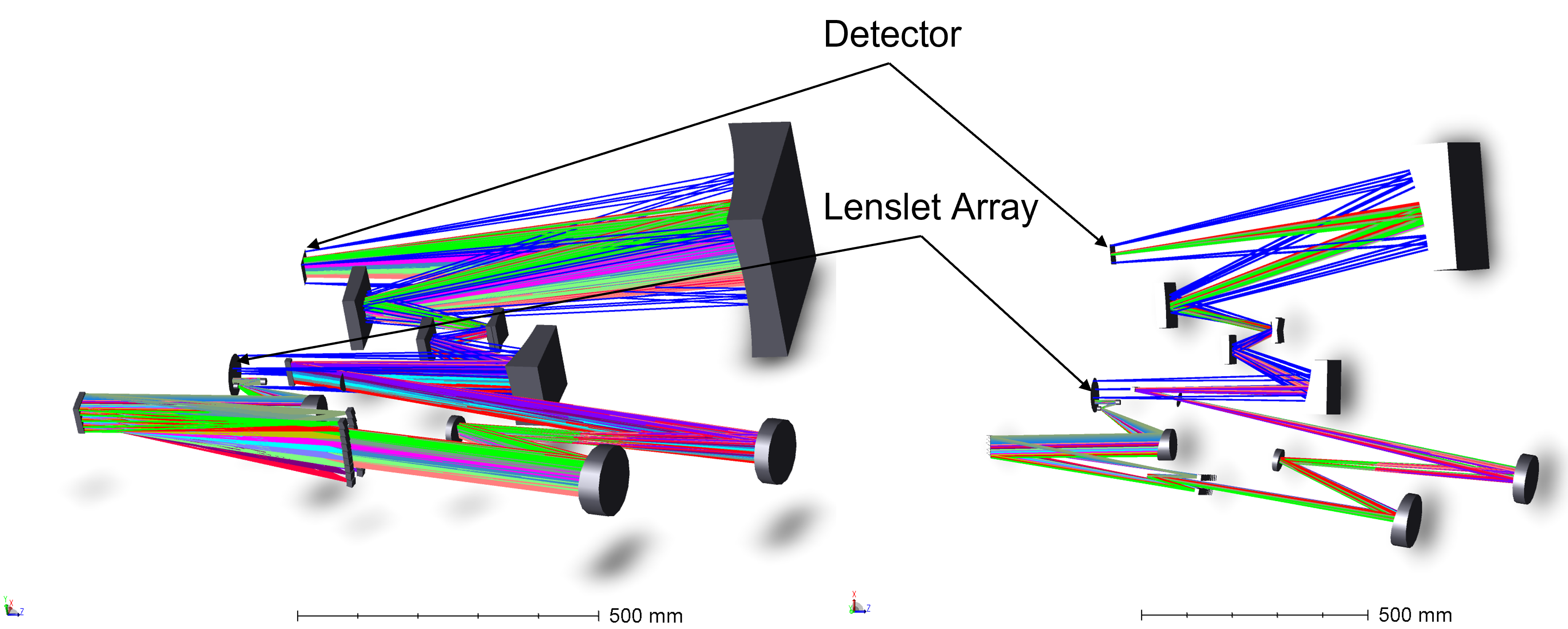}
    \caption{Optical design of the spectrograph and slenslit in Zemax.
        The blue beams are the low-resolution mode; each slice's optical beams are a single color.
        Only one wavelength is shown for each mode.
        During operation, only one mode (low or mid) will pass light to the spectrograph at a time.
        Left: Isometric view of the slenslit and spectrograph. 
        Right: Top view of the slenslit and spectrograph (rotated 15 deg about the Y axis into the paper).
    }
    \label{fig-od:slenslit-overview}
\end{figure}

The SCALES slenslit optical design (shown in Figure~\ref{fig-od:slenslit-overview}) is effectively 3 in-series focal plane-to-focal plane reimagers, and each are discussed in this subsection.
The opto-mechanical design is discussed in more detail in \S\ref{sec:mech-design}.

The first focal plane is the output micro-pupil image plane produced by the lenslet array, which produces a regular grid of micro-pupils with an f/\SI{8} beam speed. 
An input relay magnifies the beam speed from f/\SI{8} to f/\SI{64} and produces a focal plane for the slicing optics; the slicing optics geometrically rearrange the focal plane into a focal plane pseudoslit while decreasing the f-ratio to f/\SI{16}; and lastly the output relay, which returns the beam speed to f/\SI{8}. 
The real image of the pseudoslit is designed to be confocal with the lenslet array’s pupil image plane, which acts as the object for the spectrograph optics. 
Both the input and output relays are reflective TMAs. 

A mode selector mechanism blocks the primary lenslet array when using the slenslit, also hosts a return fold mirror, which feeds the slenslit output into the spectrograph. 
The disperser mechanism allows us to put gratings for various bandpasses and spectral resolutions in the optical train.

The optical prescription of the input and output relays is given in Table~\ref{tab-od:slenslit-relays-optical-prescription}, and the mechanical aspects of the design are discussed in more detail in \S\ref{sec:mech-design}.

\begin{table}[tp]
    \centering
    \caption{Slenslit input and output relay prescription.}
    \label{tab-od:slenslit-relays-optical-prescription}

\begin{tabular}{p{1.5cm} | c c c c c c}
\multirow{3}{*}{\textbf{Element}} & \textbf{Diameter} & \textbf{Center} & \textbf{RoC} & \textbf{Conic} & \textbf{Off-axis} & \textbf{4\textsuperscript{th} Order} \\
    & \textbf{[mm]}     & \textbf{Thickness}    & \textbf{[mm]}  &    \textbf{Constant}   & \textbf{Distance} & \textbf{Term} \\
    &   &  \textbf{[mm]}  &   &  &  \textbf{[mm]}   & \\ \midrule
\textbf{IRM1} & \num{8.5}  & 8  & \num{58.143} cnc  & 0.0 & 10.5 & $-5.8418E-8$ \\
\textbf{IRM2} & \num{7.1}  & 8  & \num{54.107} cvx  & 0.0 & 3.2 & - \\
\textbf{IRM3} & \num{50}  & 10  & \num{336.792} cnc & 0.0 &  82.9 & $1.3941E-9$\\
\textbf{ORM1} & \num{85}  &  15 & \num{863.071} cnc & -0.1422 & 125 & - \\
\textbf{ORM2} & \num{45.5}  & 15  & \num{401.686} cvx & -1.0234 & 40 & -  \\
\textbf{ORM3} & \num{99}  & 25  & \num{791.921} cnc & 0.0 & 3 & - \\ 
\bottomrule
\end{tabular}
\end{table}
\subsection{Slenslit input}
\label{subsec-od:slenslit-input}

\begin{figure}[htp]
    \centering
    \includegraphics[width=0.95\textwidth]{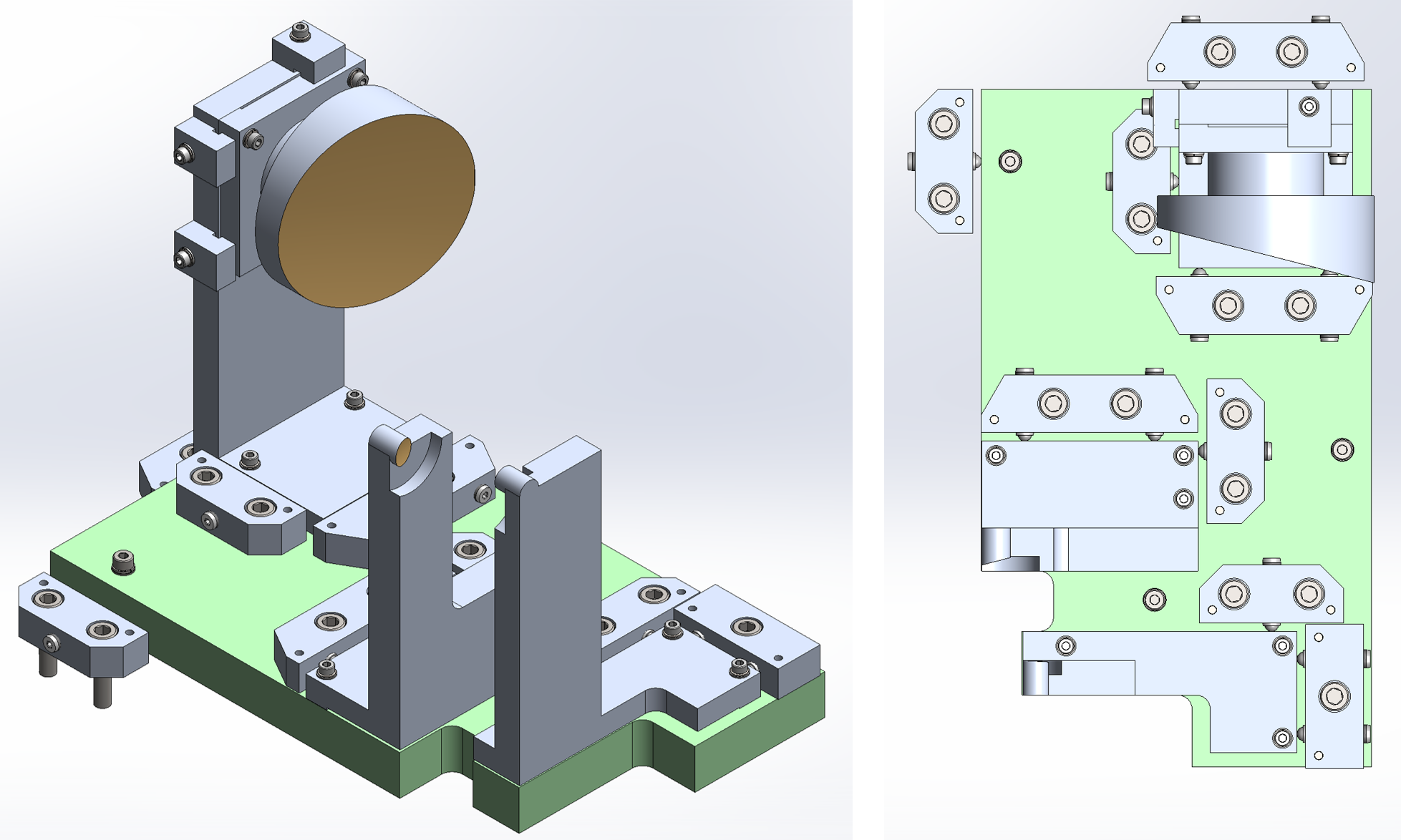}
    \caption[Slenslit input relay.]{
        Slenslit input relay.
        IRM2 is closest to the bottom in both images; IRM1 is in the middle, and IRM3 is the largest mirror at the top.
        Left: Isometric view.
        Right: Top view.
    }
    \label{fig-od:slenslit-input-relay}
\end{figure}

The input TMA relay is responsible for taking the lenslet array output (a regular grid of spots operating at f/8) and providing a magnified focal plane (operating at f/\SI{64}).  
It consists of 3 mirrors in a TMA format, and the centers of curvature are coplanar (although not co-located). 
It takes a field of view of $\sim$\SI{6.2}{ \milli\meter} $\times$ $\sim$\SI{6.0}{\milli\meter}, ($18 \times 17$ lenslets, at \SI{0.341}{ \milli\meter} pitch) and outputs a focal plane $8\times$ larger onto the slicer optics. 
The mirrors are designed to not vignette the low-resolution optical beam path. Figure~\ref{fig-od:slenslit-input-relay} shows the input relay on its sub-bench.

\subsection{Slenslit slicing optics}
\label{subsec-od:slenslit-slicing-optics}
\begin{figure}[htp]
    \centering
    \includegraphics[width=0.95\textwidth]{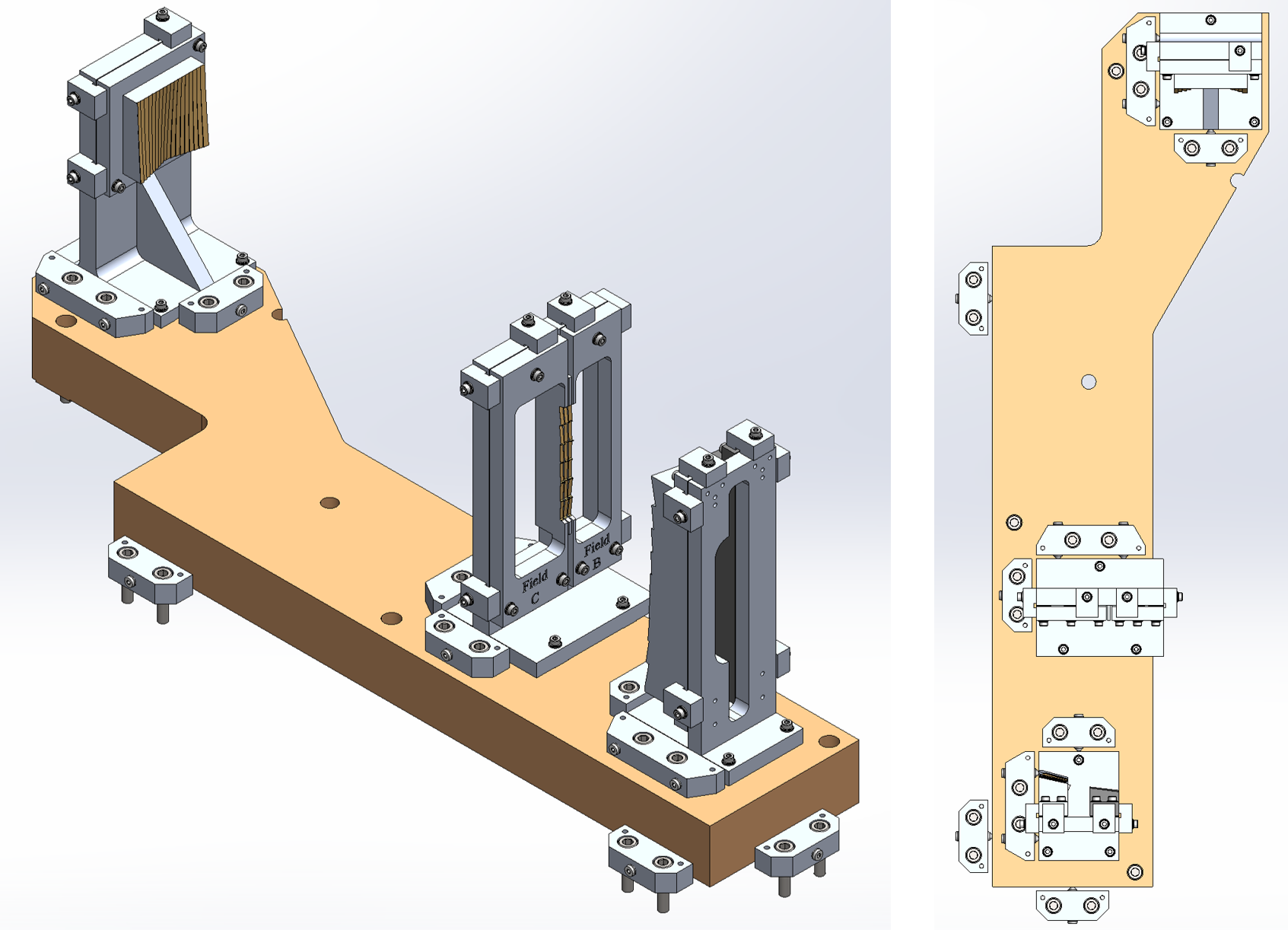}
    \caption[Slenslit slicing optics.]{
        Slenslit slicing optics.
        The slicer mirror is at the top, with the field mirrors in the middle in both images.
        Left: Isometric view.
        Right: Top view.
    }
    \label{fig-od:slenslit-slicer-optics}
\end{figure}

The slicing optics are grouped into 3 sets of mirrors, described in detail below and shown in Figure~\ref{fig-od:slenslit-slicer-optics}. 
The slicer mirror lives at the f/\SI{64} focal plane outputted by the input TMA relay. 
For ease of manufacturing, all of the slicing optics are spherical.
The slices each have optical power to produce a pupil plane roughly \SI{470}{ \milli\meter} away. 
The pupil mirrors each live at the location of the pupil image produced by the slicer, and are arranged in 2 columns. 
They, in turn, produce a focal plane inhabited by the field mirrors. 
The slicer, pupil, and field mirrors all have optical power; the field mirrors are powered in order to correct any atelecentricities introduced by the slicer and pupil mirrors. 
The output of the slicing optics produces an f/\SI{16} beam (e.g., $4\times$ demagnification).
The field mirror's three super-columns are slightly offset in the vertical direction, allowing each lenslet pupil image to be dispersed in the horizontal direction across the full width of the detector.

A subtlety of diffraction-limited image slicers is that the pupils of each slice are diffracted.
In order to prevent PSF-broadening, the pupil mirrors are slightly elongated along the axis perpendicular to the slice's long axis, which captures the diffracted energy.
The pupil mirrors could be made with circular apertures, but it's possible to more tightly pack them by elongating the mirrors along one axis (this approach was used by the FRIDA slicing optics as well~\cite{CuevasFRIDA2014}).
This also reduces higher-order aberrations by keeping the angles of incidence and reflection smaller for the most marginal mirrors.

The slicing optics’ typical optical prescription ranges are described in Table~\ref{tab-od:slenslit-slicing-optics-optical-prescription}; note that each set of slice, pupil, and field mirrors have unique tips, tilts, and radii of curvature, although all mirrors are spheres.

\begin{table}[tp]
    \centering
    \caption{Typical slenslit slicer prescription.}
    \label{tab-od:slenslit-slicing-optics-optical-prescription}

\begin{tabular}{p{2cm} | c c c}
\multirow{2}{*}{\textbf{Element}} & \textbf{Length } & \textbf{Width} & \textbf{RoC} \\
    & \textbf{[mm]}     & \textbf{[mm]}    & \textbf{[mm]}  \\ \midrule
\textbf{slicer} & 50  & 2.5  & 910 cnc  \\
\textbf{pupil} & 13.5 & 18  & 210 cnc \\
\textbf{field} & 11.6  & 1  & 250 cnc   \\ \bottomrule
\end{tabular}
\end{table}

\subsection{Slenslit output}
\label{subsec-od:slenslit-output}
\begin{figure}[htp]
    \centering
    \includegraphics[width=0.95\textwidth]{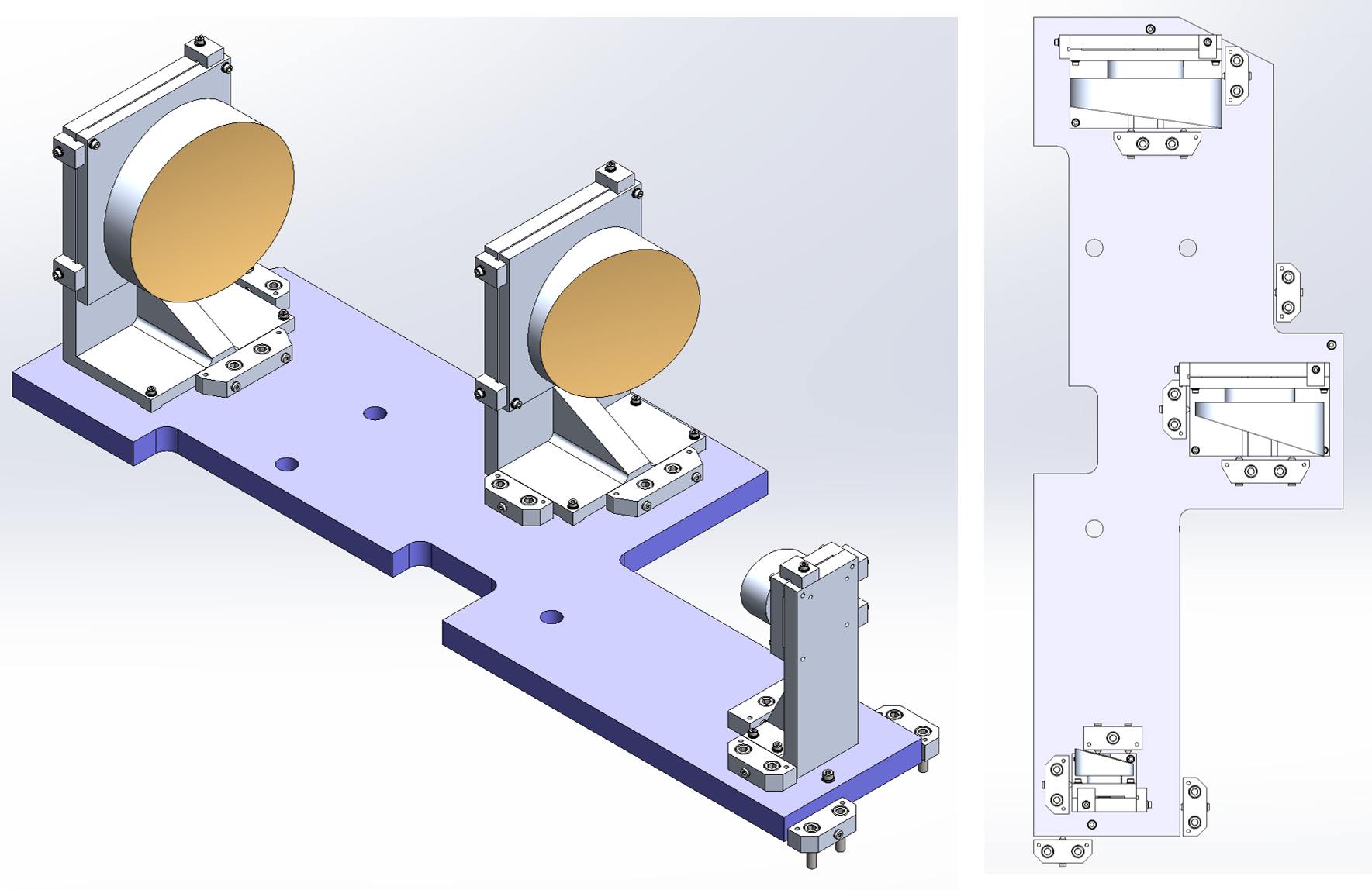}
    \caption[Slenslit output relay.]{
        Slenslit output relay.
        ORM1 is in the middle, with ORM2 at the bottom and ORM3 at the top of both images.
        Left: Isometric view.
        Right: Top view.
    }
    \label{fig-od:slenslit-output-relay}
\end{figure}

The output TMA relay is responsible for taking the output of the slicer (the three staggered super-columns) and providing a de-magnified focal plane (operating at f/\SI{8}) that is confocal with the lenslet array’s pupil image plane.  
It consists of 3 mirrors in a TMA format, and the centers of curvature are co-planar and co-linear (although not co-located); the optical prescription is shown in Table~\ref{tab-od:slenslit-relays-optical-prescription}. 
Figure~\ref{fig-od:slenslit-output-relay} shows the output relay on its sub-bench.
This allows us to use the same spectrograph optics as the low-resolution mode of SCALES.

\section{MECHANICAL DESIGN OF THE SLENSLIT}
\label{sec:mech-design}
\begin{figure}[htp]
    \centering
    \includegraphics[width=0.95\textwidth]{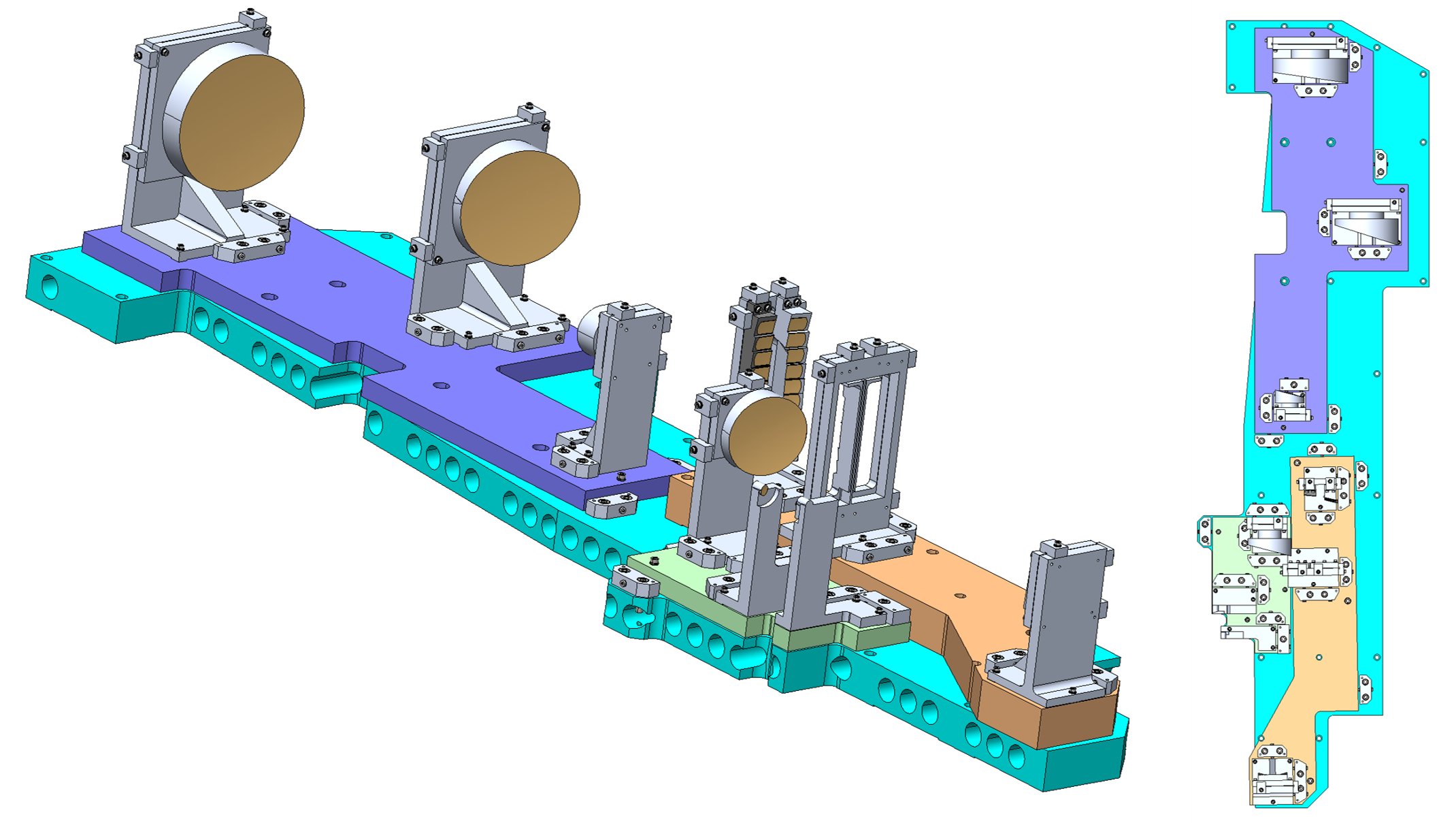}
    \caption[SCALES slenslit.]{
        SCALES slenslit. 
        All components except fasteners are machined from RSA (mirrors) or regular (brackets and benches) \SI{6061}-T\SI{6} aluminum.
        The teal bench is the main slenslit bench, with the green, tan, and purple sub-benches being the input, slicing, and output sub-benches, respectively.
        Left: Isometric view of the slenslit. 
        Right: Top view of the slenslit assembly.
    }
    \label{fig-od:slenslit-mech-overview}
\end{figure}

Within the slicer module, there are three (\num{3}) mirrors in the input TMA relay, fifty-two (\num{52}) mirrors in the slicing optics spread over six (\num{6}) substrates, three (\num{3}) mirrors in the output TMA relay, and one (\num{1}) flat fold return mirror.
In total, the slenslit has \num{59} mirrors over \num{13} substrates; however, each micro-pupil `sees' only \num{10} reflections as it flashes through the `scenic byway' on its way to the SCALES spectrograph.
Figure~\ref{fig-od:slenslit-mech-overview} shows the opto-mechanical design minus the flat fold return mirror.

We have modeled all mirrors with realistic toolpaths in SolidWorks to ensure no mirrors cut into neighboring mirrors (i.e., fratricide) and provide appropriate toolpath reliefs (i.e., no infinitely sharp corners for cutting features).  
Each mirror's geometry is defined in SolidWorks by using the Zemax optical prescription (e.g., center of curvature, radius of curvature, mirror aperture vertex, etc.). 
To simplify the SolidWorks model, all surfaces are modeled as spheres -- before fabrication, a point cloud will be generated with the conic and aspheric terms added in where appropriate.
The substrates start out as raw metal blanks, with mirror features added sequentially.
Location features, such as bolt pads on the rear surfaces, are included in the mechanical design; this means the design inherently carries with it features that are defined relative to the mirror's geometry (e.g., center of curvature, aperture center, centering pin, etc.).

The opto-mechanical design calls for each of the three sub-assemblies (input relay, slicing optics, output relay) to be mounted on their own sub-benches which in turn are mounted to a larger sub-bench mounted to the main optical bench. 
This particular design approach allows for each of the three sub-assemblies (input relay, slicing optics, output relay) to be aligned separately before aligning the sub-assemblies to each other.

Further, each mirror mount on its sub-bench is initially defined by use of 6061-T6 `nudger blocks,' each of which carries one or two fine-pitched ball-end screws that provide three points of contact to the mount (this constrains the position of the mount on the plane of the sub-bench).
The ball-end screws can push the mounts around during alignment, and are designed to provide flexible, highly repeatable location information if the mounts need to be removed for any reason.
Before cooling down to cryogenic temperatures, the nudger blocks will be replaced with three (\num{3}) 6061-T6 eccentric cams that will rotate about the bolt hole formerly used by the nudger blocks and come into contact with the mount before being locked down with its own bolt.
The cams will sit in a precision counterbore on the sub-bench centered at the bolthole.
This approach ensures that the location of the mounts on the bench are defined with as little over-constraint as possible.

In order to produce optics with low tolerance stack-up and to simplify alignment, the mirrors share substrates whenever possible. 
The slicer is carved from a monolithic block of aluminum, while the two columns of pupil mirrors are fabricated from two substrates while sharing a mounting bracket. 
The field mirrors are broken up into three substrates, with one substrate acting as the mounting bracket for the other two.

Additionally the use of precision shims, attached on the tops and sides of the mounting brackets with one bolt each, allow for quick and simple metrology of reference surfaces of the mirror substrates. 
As a proof of concept for this approach, we designed a benchtop prototype slenslit slicer that was fabricated at Durham Precision Optics; it is described in a previous Proceedings~\cite{StelterColorsToChemistry2021}.
Note that the precision shims are used to define the positions of the mirror substrates on their brackets.
The mirror surfaces are well-defined in relation to the substrates during the SPDT process, so the shims are an excellent way to provide location information without unduly over-constraining the substrates.

\section{EXPECTED PERFORMANCE}
\label{sec:performance}
The mid-resolution mode offers, for the first time, multiple spaxels with spectral resolution of $\sim3000-6000$.
Figure~\ref{fig:resolution-by-filter} shows the instantaneous spectral resolution for each filter and IFS mode.
While other instruments such as KPIC~\cite{MawetKPIC2016} have access to much higher spectral resolution, they use single-mode fibers that precludes having more than a few spaxels.
In part, this is a trade-off between higher spectral resolution and the number of spatial units, as well as depending on the single-mode fiber to help with speckle suppression.

\begin{figure}[htp]
    \centering
    \includegraphics[width=0.95\textwidth]{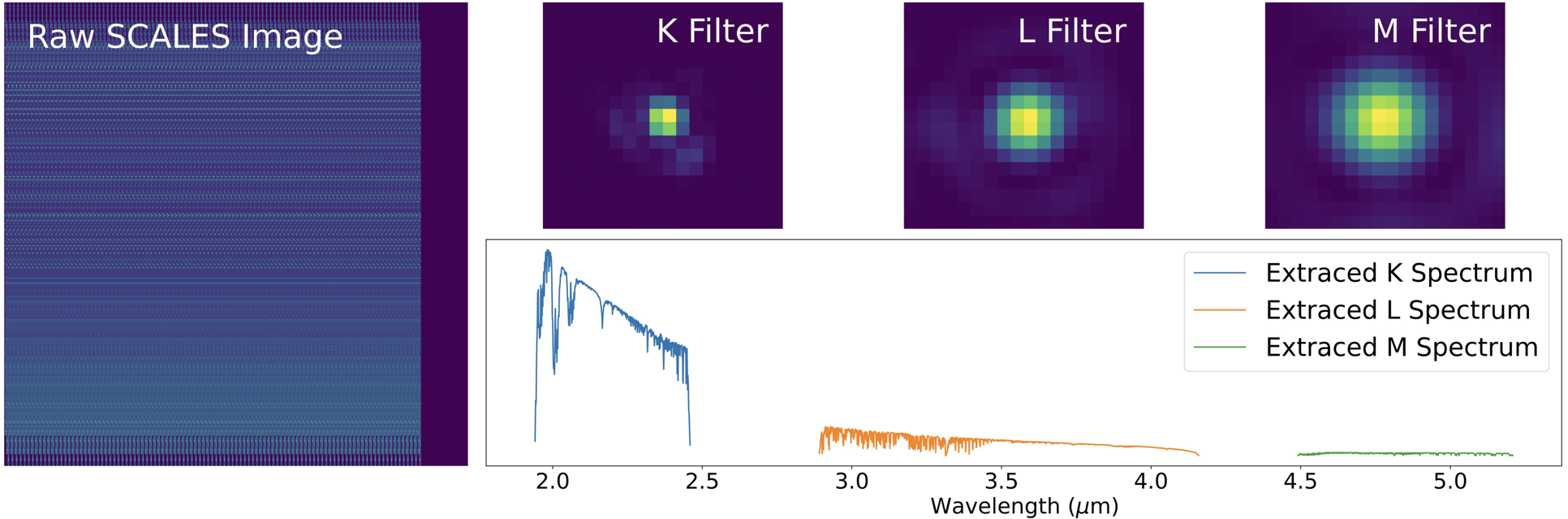}
    \caption{A simulated A0 star mid-resolution spectrum, with astrophysical and instrumental injected noise.
    \textit{Left}: the raw SCALES image.
    \textit{Right}: The extracted spectrum, with an image for each bandpass included above.
    }
    \label{fig:A0-spectrum}
\end{figure}

An example of a simulated A0 stellar spectrum is shown in Figure~\ref{fig:A0-spectrum}. 
Our simulations use the SCALES pipeline and simulator, \texttt{scalessim} (Briesemeister et al, in prep), which includes Fresnel diffraction effects from the lenslet array's lenslet and pinhole apertures.
Noise terms from atmospheric and instrumental effects are also included, but do not yet include realistic telluric correction errors.
\texttt{scalessim} is available on GitHub (\url{https://github.com/scalessim/scalessim}).

\begin{figure}[htp]
    \centering
    \includegraphics[width=\textwidth]{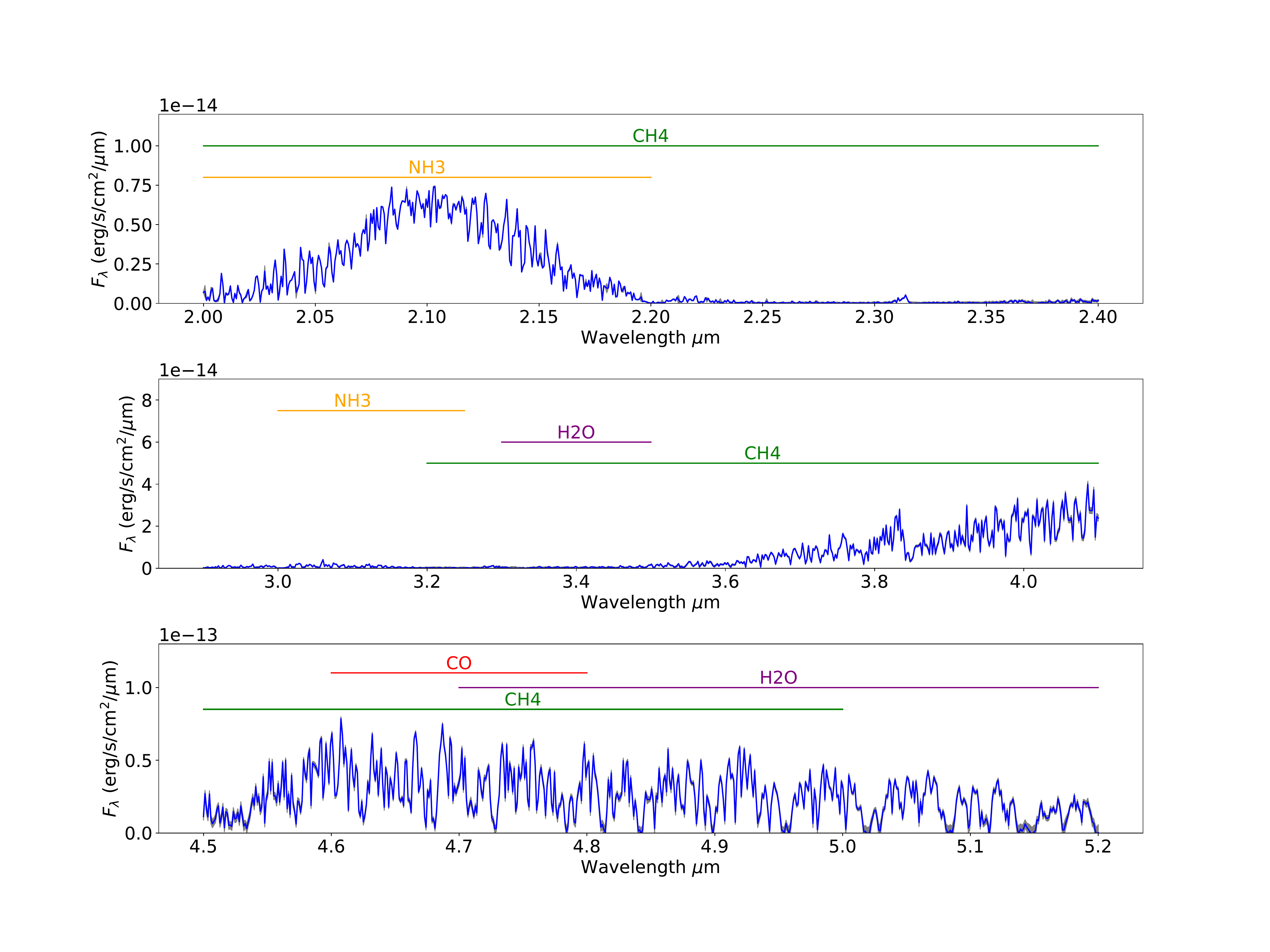}
    \caption[Simulated Gaia planet spectroscopy with SCALES.]{
        Simulated K band (top), L band (middle), and M band (bottom) medium resolution mode SCALES observations of a 500 K planet at a distance of 15 pc. 
        The assumed integration time is 10 hours, and the extracted spectrum is shown in blue with error bars overlaid in grey. SCALES' wavelength coverage will lead to constraints on molecules such as H$_2$O, CO, CH$_4$, NH$_3$ (see horizontal bars), which can be used to calculate properties such as metallicities and C/O ratios. 
        Currently, the simulations include sky/telescope background noise, but don't yet include realistic telluric correction errors, which will matter (particularly at M band).
    }
    \label{fig:exoplanet-spectrum-500K}
\end{figure}

\begin{figure}[htp]
    \centering
    \includegraphics[width=\textwidth]{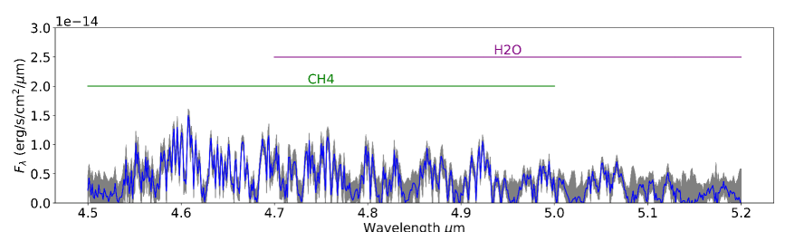}
    \caption{Simulated M band medium resolution mode SCALES observation of a \SI{300}{\kelvin} planet at a distance of \num{15} pc.
        The assumed integration time is 25 hours, and the extracted spectrum is shown in blue with error bars overlaid in grey. 
        Note this spectrum is noisier than the \SI{500}{\kelvin} planet.
        Currently, the simulations include sky/telescope background noise, but don't yet include realistic telluric correction errors, which will matter (particularly at M band).
    }
    \label{fig:exoplanet-spectrum-300K}
\end{figure}

The mid-resolution mode will allow for characterization of exo-atmospheres by measuring chemical abundances such as CO, CH\textsubscript{4}, H\textsubscript{2}O, NH\textsubscript{3}, and others.
A simulated K, L, and M band spectrum of an \SI{500}{\kelvin} exoplanet at a distance of \SI{15}{parsecs} with a \num{10} hour exposure time is shown in Figure~\ref{fig:exoplanet-spectrum-500K}.
This is significantly colder than 51 Eri b, the coldest directly-imaged exoplanet to date~\cite{macintosh201551erib} with a temperature of \SIrange{700}{750}{\kelvin} (the temperature range is due to a degeneracy in fitting cloudless vs cloudy models to the data).
A second simulated spectrum in M band of an even colder exoplanet is shown in Figure~\ref{fig:exoplanet-spectrum-300K}.
This simulated exoplanet is also at a distance of \num{15} pc and has a temperature of \SI{300}{\kelvin}.

\section{CONCLUSION}
\label{sec:conclusion}
We have presented the opto-mechanical design of the SCALES slenslit optics, and provided examples of its expected performance at medium spectral resolution in comparison to the low-resolution IFS mode.
The slenslit (a combination of lenslet array and image slicer) opens up unprecedented spatially-resolved, high-contrast, mid-resolution IFU exoplanet spectroscopy, and will further our understanding of warm and cold exoplanetary atmospheres.
SCALES is in its final design phase and we expect to be on-sky in late 2025.

\acknowledgments 
We are grateful to the Heising-Simons Foundation and the Mt. Cuba Astronomical Foundation for their generous support of our efforts.

\bibliography{report} 
\bibliographystyle{spiebib} 

\end{document}